\documentclass[conference]{IEEEtran}
\usepackage{booktabs}
\usepackage{mathrsfs}
\usepackage{array}
\usepackage{graphicx}
\usepackage{latexsym}
\usepackage{amsmath}
\usepackage{amsthm}
\usepackage{color}
\usepackage{amsfonts}
\usepackage{subfigure}
\usepackage{dsfont}
\usepackage{epstopdf}
\usepackage{threeparttable}
\usepackage{multirow}
\usepackage{enumerate}
\usepackage{epsfig}
\usepackage{bm}
\usepackage{cite}
\usepackage{algorithmic}
\usepackage{textcomp}
\usepackage{graphicx}
\usepackage{latexsym}
\usepackage{amsmath}
\usepackage{amsthm}
\usepackage{makecell,rotating,multirow,diagbox}
\usepackage{amsfonts}
\usepackage{subfigure}
\usepackage{dsfont}
\usepackage{epstopdf}
\usepackage{threeparttable}
\usepackage{multirow}
\usepackage{enumerate}
\usepackage{epsfig}
\usepackage{bm}
\usepackage{cite}
\usepackage{algorithmic}
\usepackage{textcomp}
\usepackage{xcolor}
\usepackage{amsmath}
\usepackage{cite}
\begin{document}
\title{Label Design-based ELM Network for Timing Synchronization in OFDM Systems with\\Nonlinear Distortion}
\author{\IEEEauthorblockN{Chaojin~Qing$^\ast$, Shuhai~Tang$^\ast$, Chuangui~Rao$^\ast$, Qing~Ye$^\ast$, Jiafan~Wang$^\dag$, and Chuan~Huang$^\ddag$}
\IEEEauthorblockA{$^\ast$School of Electrical Engineering and Electronic Information,
Xihua University, Chengdu, 610039, China\\
$^\dag$Synopsys Inc., 2025 NE Cornelius Pass Rd, Hillsboro, OR 97124, USA\\
$^\ddag$School of Science and Engineering, Chinese University of Hong Kong, Shenzhen, 518172, China\\
Email: $^\ast$qingchj@mail.xhu.edu.cn, $^\dag$jifanw@gmail.com, $^\ddag$huangchuan@cuhk.edu.cn}}
\maketitle
\begin{abstract}
Due to the nonlinear distortion in Orthogonal frequency division multiplexing (OFDM) systems, the timing synchronization (TS) performance is inevitably degraded at the receiver. To relieve this issue, an extreme learning machine (ELM)-based network with a novel learning label is proposed to the TS of OFDM system in our work and increases the possibility of symbol timing offset (STO) estimation residing in inter-symbol interference (ISI)-free region. Especially, by exploiting the prior information of the ISI-free region, two types of learning labels are developed to facilitate the ELM-based TS network. With designed learning labels, a timing-processing by classic TS scheme is first executed to capture the coarse timing metric (TM) and then followed by an ELM network to refine the TM. According to experiments and analysis, our scheme shows its effectiveness in the improvement of TS performance and reveals its generalization performance in different training and testing channel scenarios.
\end{abstract}
\IEEEpeerreviewmaketitle

\section{Introduction}
Orthogonal frequency division multiplexing (OFDM) system is pervasively applied in modern communication systems, such as wireless local area networks (WLAN) \cite{ref_DWIFI} and the upcoming fifth-generation (5G) wireless communication system~\cite{Ref_5G}. In the OFDM system, the overall performance heavily relies on the process of timing synchronization (TS). Thus, in the past two decades, lots of classic TS schemes emerged for OFDM systems. However, a large number of non-linear devices or blocks usually exist in the OFDM system, e.g., high power amplifier (HPA), digital to analog converter (DAC), thereby causing nonlinear distortion \cite{ref_distort} and degrading the receiver’s TS performance. To this end, the nonlinear distortion needs to be considered in the OFDM system design. Owing to the lack of consideration for nonlinear distortion, the existing TS schemes are facing great challenges.

Due to the prominent ability to cope with nonlinear distortion, machine learning has drawn considerable attention in recent years \cite{ref_elmch,ref_dlch}. Machine learning, especially deep learning (DL), has been widely applied in wireless communication systems \cite{ref_dlch,ref_dlcode,ref_dlcsi,ref_cemm,ref_dlcede}, e.g., signal detection \cite{ref_dlch}, precoding \cite{ref_dlcode}, channel state information (CSI) feedback \cite{ref_dlcsi}, and channel estimation \cite{ref_cemm,ref_dlcede}. However, there are limited DL-based proposals for the TS scheme subject to nonlinear distortion. Compared with DL-based approaches, the extreme learning machine (ELM) is raised in~\cite{ref_GBelm}, which is a single hidden layer feed-forward neural network with no requirement of gradient back-propagation. Relative to DL-based networks, the ELM network presents many advantages, such as less time-consuming for network training and good generalization performance~\cite{ref_GBelm}.

In this paper, we introduce the ELM-based network into the TS of the OFDM system to improve the adaptability of the existing classic TS scheme \cite{ref_sandc} against nonlinear distortion. Considering that the learning ability of a neural network is influenced by learning labels to a certain extent~\cite{ref_label} and the ELM-based network is no exception, we develop the learning label to enhance the learning ability of ELM-based networks for TS in OFDM systems.

The remainder of this paper is structured as follows. In Section II, we present the system model. The designed learning label for the ELM-based TS network is proposed in Section III and illustrated in Section IV. Numerical results and analysis are presented in Section V, and Section VI concludes our work.
\begin{figure}[h]
\centering
\includegraphics[scale=0.36]{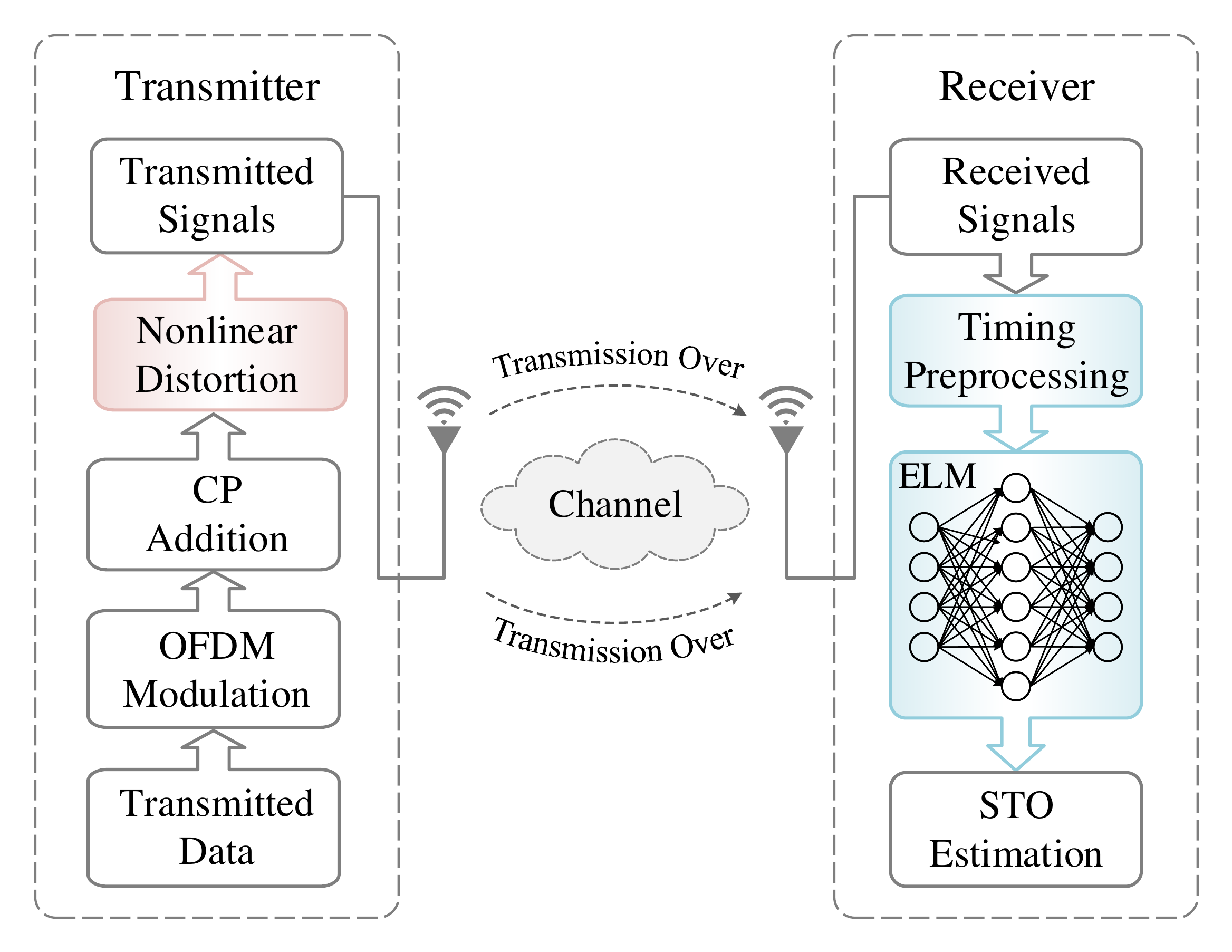}
\caption{System model.}
\label{figSys}
\end{figure}

\section{System Model}
Considering a system model in Fig.~\ref{figSys}, the transmitter encounters nonlinear distortion and the receiver is combined by the classic synchronizer and ELM network. Supposing $N$ sub-carriers for each OFDM symbol, the $n$th received sample $r\left( n \right)$ in time domain can be represented by
\begin{equation}\label{EQrxs}
{r\left( n \right) = \sum\limits_{l = 0}^{L - 1} {h\left( l \right)\tilde s\left( {n -\theta - l} \right)} e^{ {j\left(\frac{{2\pi n\upsilon }}{N} + \phi\right) }} + w\left( n \right)},
\end{equation}
where $\theta$, $\nu$, and $\phi$ stand for the unknown STO, carrier frequency offset, and initial phase, respectively. $w\left( n \right)$ is the complex additive white Gaussian noise (AWGN) with zero-mean and variance $\sigma ^ 2$, i.e., $w\left( n \right)\sim \mathcal{CN} \left(  0, \sigma ^ 2 \right)$. $h\left( l \right)$, $l=0,2,\cdots,L-1$, represents the channel impulse response (CIR) with a memory of $L$-length samples. $\widetilde{s}\left( n \right)$ is the OFDM signal suffered from nonlinear distortion at transmitted, i.e.,
\begin{equation}\label{EQdis}
\widetilde{s}\left( n \right) = {f_{{\rm{dis}}}}\{ {s\left( n \right)} \},
\end{equation}
where $f_{\mathrm{dis}}\{\cdot\}$ denotes an universal process for nonlinear distortion, e.g., PA, DAC with hardware imperfections. $s\left(n\right)$ is the undistorted OFDM signal, which can be expressed as
\begin{equation}\label{EQtxs}
{s\left( n \right) = \sum\limits_{k = 0}^{N - 1} {d\left( k \right){e^{j\frac{{2\pi nk}}{N}}}} },
\end{equation}
where $d\left( k \right)$, $k=0,1,\cdots,N-1$, is the $k$th frequency-domain symbol, which satisfies $\mathbf{E}\{|d\left( k \right)|^2\}= \sigma _d ^2$ with $ \sigma _d$ being constant. In (\ref{EQrxs})-(\ref{EQtxs}), the discrete-time index $n$ satisfies $ -N_g \leq n \leq N-1$, where $N_g$ stands for the length of cyclic prefix (CP). Without loss of generality, $N_g > L$ is assumed.

With the received samples in (\ref{EQrxs}), an classic TS synchronizer is first performed to calculate the TM, and then an ELM-based network is applied to alleviate the impacts of nonlinear distortion and refine the TM for TS.

\section{Learning-label design}
In machine learning, a good label usually facilitates the learning ability of the neural network \cite{ref_label}. To enhance the learning ability of ELM-based network, an effective label design needs to be concerned.

For expression convenience, a general label-vector form, denoted as $\mathbf{T}\in \mathbb{R} ^{N_{d}\times1}$, is employed by using a time-indexed sequence, i.e.,
\begin{equation}\label{EQLform}
{\mathbf{T} = {\left[ T_0, T_1, \cdots,T_{n}, \cdots, T_{N_d - 1} \right]^T}},
\end{equation}
where $N_d$ is an observed window length within TM, and $T_n$, $n = 0,1,\cdots,N_d-1$, corresponds to the $n$th label value in a window of $N_d$ samples of TM.

\subsection{Learning Label Using One-hot encoding}
In \cite{ref_fsELM}, a label using one-hot encoding is considered.
\begin{equation}\label{EQLone}
{{T_n} = \left\{ {\begin{array}{*{20}{l}}
{1,}&{n = { \theta + N_{g} +1}}\\
{0,}&{\textrm{others}}
\end{array}} \right.},
\end{equation}
however, the label in (\ref{EQLone}) is vulnerable to be interfered with multi-path, noise, and nonlinear distortion, etc. Also, this label lacks tolerance in timing errors, and thus degrades the TS performance of OFDM system.

\subsection{Exploiting the midpoint of ISI-free region}
For the cases where the length of CP is larger than the length of multi-path, the midpoint of ISI-free region is relatively stable. Therefore, we exploit the superiority of the midpoint of ISI-free region, making the STO estimation reside in the ISI-free region. The label form can be expressed by
\begin{equation}\label{EQL1}
{{T_n} = \left\{ {\begin{array}{*{20}{l}}
{1,}&{n = { n_c}}\\
{0,}&{\textrm{others}}
\end{array}} \right.},
\end{equation}
where $n_c$ denotes the time index of the midpoint of ISI-free region, i.e., $n_c = \theta+\lfloor ( N_g+L+1)/2\rfloor$. In this paper, the label in (\ref{EQL1}) is called as \textit{midpoint-based learning label} and denoted by $\mathbf{T}_{\textrm{mid}}$, which is expressed as
\begin{equation}\label{EQL1form}
{{\mathbf{T}_\textrm{mid}} = {\left[ { {\mathbf{0}}_{  {n_c} - 1},1, {\mathbf{0}}_{{N_d} -  {n_c}}} \right]^T}},
\end{equation}
where $\mathbf{0}$ denotes an all-zeros row vector.
For the label $\mathbf{T}_{\textrm{mid}}$, its performance of STO estimation is further improved compared with the label in \cite{ref_fsELM}. Even so, only the midpoint of ISI-free region within TM is considered, while the full use of prior information of ISI-free region is not being considered. This impels us to further improve the learning-label design.

\subsection{Exploiting the prior information of ISI-free region}
Since the TS of OFDM system only requires the STO estimation to determine one of valid time indexes in the ISI-free region, the full use of prior information of ISI-free region needs to be considered in the label design. By taking the full prior information of ISI-free region into account, we set the values of $T_{n}$ inside ISI-free region as 1, while setting the values of $T_n$ outside ISI-free region as 0, i.e.,
\begin{equation}\label{EQLdesign}
{{T_n} = \left\{ {\begin{array}{*{20}{l}}
{1,}&{\theta  + L \le n \le \theta  + {N_g} + 1}\\
{0,}&{\textrm{others}}
\end{array}} \right.}.
\end{equation}
In this paper, the label in (\ref{EQLdesign}) is referred as \textit{{ISI-free-based learning label}} and denoted by $\mathbf{T}_{\textrm{ISI-free}}$, which is constructed as
\begin{equation}\label{EQLcons}
{\mathbf{T}_{\textrm{ISI-free}} = {\left[ { {\mathbf{0}}_{\theta  + L - 1}, {\mathbf{1}}_{{N_g} - L + 2}, {\mathbf{0}}_{{N_d} - {N_g} - \theta -1}} \right]^T}},
\end{equation}
where $\mathbf{1}$ denotes an all-ones row vector. Since the $\mathbf{T}_{\textrm{ISI-free}}$ enhances the robustness against interference and enlarges the tolerance for timing error, the output of network could be regarded as a refined TM. Therefore, training ELM using $\mathbf{T}_{\textrm{ISI-free}}$ could guide the current STO estimation into the correct timing range of ISI-free.

\section{ELM-based TS Scheme}
In this section, we employ ELM network combined with developed learning label to tackle the TS problem.
\subsection{Timing Preprocessing}
To improve the efficiency of ELM learning, the classic synchronizer is used as a timing preprocessing to execute the \textit{knowledge discovery}. By using \textit{Schmidl's scheme} in~\cite{ref_sandc}, the timing preprocessing is performed to coarsely capture the TM. The time preprocessing is given by
\begin{equation}\label{EQTM}
{{M}\left( d \right) = \frac{{{{\left| {{P}\left( d \right)} \right|}^{\rm{2}}}}}{{{{\left| {{R}\left( d \right)} \right|}^2}}}},
\end{equation}
where $d$ is the trial value to search the start point of OFDM symbol in a window of $N_{d}$ samples. ${P}\left( d \right)$ and ${R}\left( d \right)$ stand for the autocorrelation function and normalized function separately~\cite{ref_sandc}. Considering the observed TM with length $N_d$, the TM vector $\mathbf{g} \in \mathbb{R}^{N_d \times 1}$ can be given by ${{\mathbf{g}} = {\left[ {{M\left(0\right)},{M\left(1\right)}, \cdots,M \left(d\right), \cdots ,{M{{\left( N_d -1 \right)}}}} \right]^T}}$.

The TM contains coarse features of unknown STO and nonlinear distortion, which could be regarded as the knowledge for ELM learning. For further facilitating the training and testing of ELM network, the TM vector $\mathbf{g}$ is normalized by
\begin{equation}\label{EQ11}
{\overline {\mathbf{g}} = \frac{{\mathbf{g}}}{{{{\left\| {\mathbf{g}} \right\|}_2}}}}.
 \end{equation}
With the normalized TM vector $\overline{\mathbf{g}}$, the combination of ELM network and designed learning label is employed for the TS in OFDM systems to relieve the nonlinear distortion and refine the TM.

\subsection{Network Training and Testing}
In this subsection, we present the ELM-based TS network for OFDM system, in which the offline and online procedures are separately elaborated in the following.
\subsubsection{Off-line training specification}
In this phase, we use the ELM to learn the complex relationship of output and input, in which the training data-set $\left\{ {\left( {{\overline{\mathbf{g}}_i},{\mathbf{T}_i}} \right)} \right\}_{i = 1}^{{N_t}} \in {\mathbb{R}^{ {N_d}\times 1}}$ consists of normalized TMs $\widetilde{\mathbf{g}}_i$ and learning label $\mathbf{T}_i$. During the off-line training procedure, all the elements of input weight matrix $\mathbf{W} \in \mathbb{C} ^{\widetilde{N} \times N_d}$  and hidden bias $\mathbf{b} \in \mathbb{C} ^{\widetilde{N} \times 1}$ are randomly selected~\cite{ref_GBelm}, in which $\widetilde{N}$ is the number of hidden neurons.

In ELM, the $i$th hidden layer output $\mathbf{H}_i$ is presented as
\begin{equation}\label{EQ_Hiddenout}
{{\bf{H}}_i} = \sigma \left( {{\bf{W\widetilde{g}}}_i + {\bf{b}}} \right),
\end{equation}
where $\sigma\left(\cdot\right)$ is  hyperbolic tangent (tanh) activation function. By collecting $N_{t}$ samples of $\mathbf{H}_{i}$, a hidden layer output matrix is constructed as $\mathbf{H}=[\mathbf{H}_{1},\mathbf{H}_{2},\cdots,\mathbf{H}_{i},\cdots,\mathbf{H}_{N_{t}}]$. Correspondingly, a label matrix $\widetilde{\mathbf{T}} \in \mathbb{R} ^{N_{d} \times N_{t}}$ is obtained by loading $N_{t}$ samples of $\mathbf{T}_{i}$, i.e., $\mathbf{\widetilde{T}}=[\mathbf{T}_{1},\mathbf{T}_{2},\cdots,\mathbf{T}_{i},\cdots,\mathbf{T}_{N_{t}}]$.

With $\mathbf{H}$ and $\mathbf{\widetilde{T}}$, an optimized output weight matrix $\mathbf{\Upsilon} \in \mathbb{C}^{N_{d} \times \widetilde{N}}$ is obtained by
\begin{equation}\label{EQWout}
{\mathbf{\Upsilon}  = \widetilde{\mathbf{T}}{\mathbf{H}^\dag }}.
\end{equation}
So far the offline training is completed, and then the online deployment could be implemented according to $\mathbf{W}$, $\mathbf{b}$, and $\mathbf{\Upsilon}$.
\subsubsection{Online running}
In this phase of ELM testing, the received signals $r\left(n\right)$ are first inputted to the classic synchronizer for knowledge discovery. Then, we feed normalized TM $\widetilde{\mathbf{g}}$ into the trained ELM network to obtain refined TM $\mathbf{O}\in \mathbb{R}^{N_d \times 1}$, i.e.,
\begin{equation}\label{EQ_RTM}
{\mathbf{O} = \mathbf{\Upsilon}  \cdot \sigma \left( {\mathbf{W}\widetilde{\mathbf{g}} + \mathbf{b}} \right)}.
\end{equation}
By expressing $\mathbf{O}$ as $\{O_d\}^{N_d-1}_{d=0}$, the STO estimation is obtained by $\hat \theta  = \arg {\max _{0 \le d \le {N_d} - 1}} {{\left| {{O_d}} \right|^2}} $. It seems that  straightforward since the complicated work has been released to the training phase of ELM-based TS. Besides, the known network parameters (i.e., $\mathbf{W}$, $\mathbf{b}$, and $\mathbf{\Upsilon}$) can accelerate the processing of ELM-based TS networks according to parallel processing mode, which leads to a low processing delay.
\section{Experimental Analysis}
In this section, numerical results are provided to illustrate the performance of ELM-based TS scheme for OFDM system. The basic parameters and definitions involved in the simulations are given in the following.

\subsection{Parameter Setting}
In the simulations, the basic parameters $N=64$, $N_g=16$, $N_d=2\left(N+N_g\right)=160$, $\widetilde{N}=8\left(N+N_g\right)=640$, $N_t=2^{17}$ are considered, respectively. The signal-to-noise ratio (SNR) in decibel (dB) is defined as $\mathrm{SNR} \left(\textrm{dB}\right)=10\log_{10}\left(\sigma^{2}_{P}/\sigma^{2}\right)$~\cite{Ref_dcsiQ}. Without loss of generality, $\sigma_P^2 = \sigma_d^2$ is considered in the simulations, i.e., the preamble and data are assigned the same transmitted power. According to \cite{ref_pr}, the error probability of TS means that probability of STO estimation falling outside of the ISI-free region, i.e., ${{\cal P}_{e,\texttt{TS}}} = {\rm{Pr}}\left\{ {\hat \theta  \notin \left[ {\theta  + L,\theta  + {N_g} + 1} \right]} \right\}$.

In the simulations, the nonlinear amplitude $A\left(r\right)$ and phase $\phi\left(r\right)$, $r=|s\left(n\right)|$, are defined as $A\left( r \right) = {{{\alpha _a}r}}/{\left({1 + {\beta _a}{r^2}}\right)}$ and $\Phi\left( r \right) = {{{\alpha _\phi }{r^2}}}/{\left({1 + {\beta _\phi }{r^2}}\right)}$ respectively, in which $\alpha_a=1.96$, $\beta_a=0.99$, ${{\alpha _\phi }}={2.53}$, and $\beta_\phi=2.82$~\cite{ref_dis}. Also, the error vector magnitude (EVM) is used to evaluate the distortion intensity~\cite{ref_evm}, i.e.,
\begin{equation}
\mathrm{EVM}\left( \%  \right) = \sqrt {\frac{{\sum {{{\left| {\tilde s\left( n \right) - {s_{\texttt{ref}}}\left( n \right)} \right|}^2}} }}{{\sum {{{\left| {{s_{\texttt{ref}}}\left( n \right)} \right|}^2}} }}},
\end{equation}
where $s_{\texttt{ref}}\left(n\right)$ denotes the $n$th reference signal amplified by HPA linearly, i.e., transmitted signal without distortion.

For expression convenience, ``{Ref\_\cite{ref_fsELM}}'', ``{Prop\_$\textrm{T}_{\textrm{mid}}$}'', and ``{Prop\_$\textrm{T}_{\textrm{ISI-free}}$}'' are employed to denote the ELM-based TS scheme with the label used in~\cite{ref_fsELM} and the label in (\ref{EQL1}), and the label in (\ref{EQLdesign}), respectively. The classic TS scheme proposed in~\cite{ref_sandc} serves as a baseline and is referred to as ``{SC\_corr}'' in the simulations. Meanwhile, we employ ``TS\_Learn'' to represent the ELM-based TS scheme without timing preprocessing, i.e., directly employing the received signal as the network input.
\begin{figure}[h]
\centering
\includegraphics[scale=0.45]{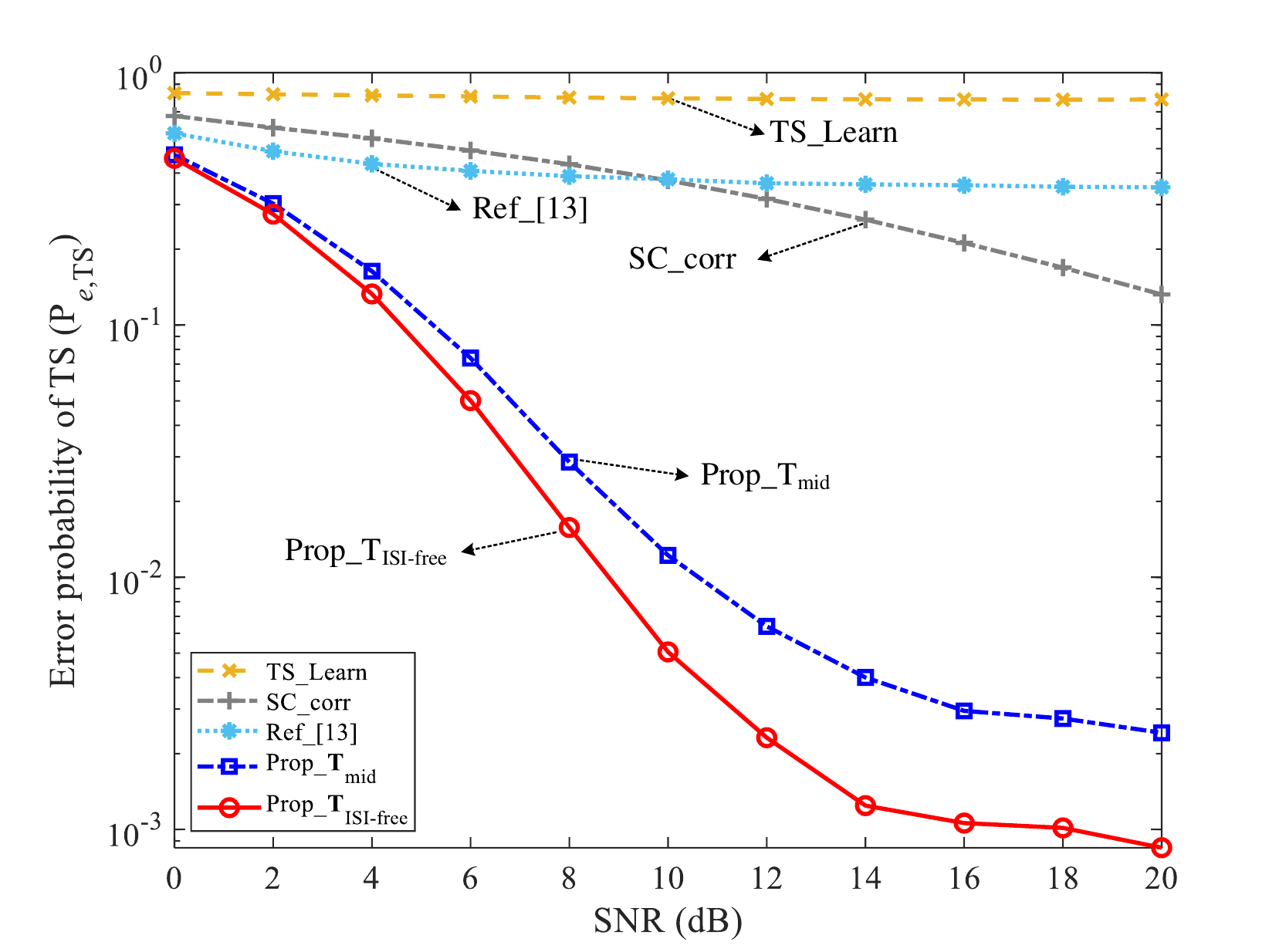}
\caption{Comparison of TS error probability, where $N = 64$, $N_g = 16$, $L =8$, and $\mathrm{EVM} = 40\%$ are considered.}
\label{Sim1}
\end{figure}
\subsection{Error probability of TS}
The effectiveness of the proposed ELM-based TS is verified in terms of error probability curves in Fig.~\ref{Sim1}, where $L =8$ and ${\mathrm{EVM}} = 40\%$ are considered. From Fig.~\ref{Sim1}, the error probability of ``TS\_Learn'' is higher than those of ``SC\_corr'', ``{Ref\_\cite{ref_fsELM}}'', ``{Prop\_$\textrm{T}_{\textrm{mid}}$}'', and ``{Prop\_$\textrm{T}_{\textrm{ISI-free}}$}'' for all the SNR values. This demonstrates that the ELM-based TS network cannot work well without timing preprocessing. Meanwhile, the error probability of ``SC\_corr'' is improved by ``{Ref\_\cite{ref_fsELM}}'' for low SNRs (e.g., $\mathrm{SNR}<10$dB), yet ``{Ref\_\cite{ref_fsELM}}'' is inapplicable due to the relatively high error probability at the high SNR region (e.g., $\mathrm{SNR}>10$dB). Nevertheless, the lower error probabilities of ``{Prop\_$\textrm{T}_{\textrm{mid}}$}'' and ``{Prop\_$\textrm{T}_{\textrm{ISI-free}}$}'' retain the feasibility for practical applications in the relatively high SNR region, e.g., $\mathcal{P}_{e,\texttt{TS}}<0.5\times10    ^{-2}$ for ``{Prop\_$\textrm{T}_{\textrm{ISI-free}}$}'' when $\mathrm{SNR}\geq12$dB. It is also worth noting that, ``{Prop\_$\textrm{T}_{\textrm{ISI-free}}$}'' reaches the smallest error probability, and this confirms the designed rationality of ``{Prop\_$\textrm{T}_{\textrm{ISI-free}}$}''. To sum up, ``{Prop\_$\textrm{T}_{\textrm{mid}}$}'' and ``{Prop\_$\textrm{T}_{\textrm{ISI-free}}$}'' possess the effective improvement in the reduction of error probability.
\begin{figure}[h]
\centering
    \subfigure[``Prop\_$\textrm{T}_{\textrm{mid}}$'']{
    \label{figL1_General}
    \includegraphics[scale=0.35]{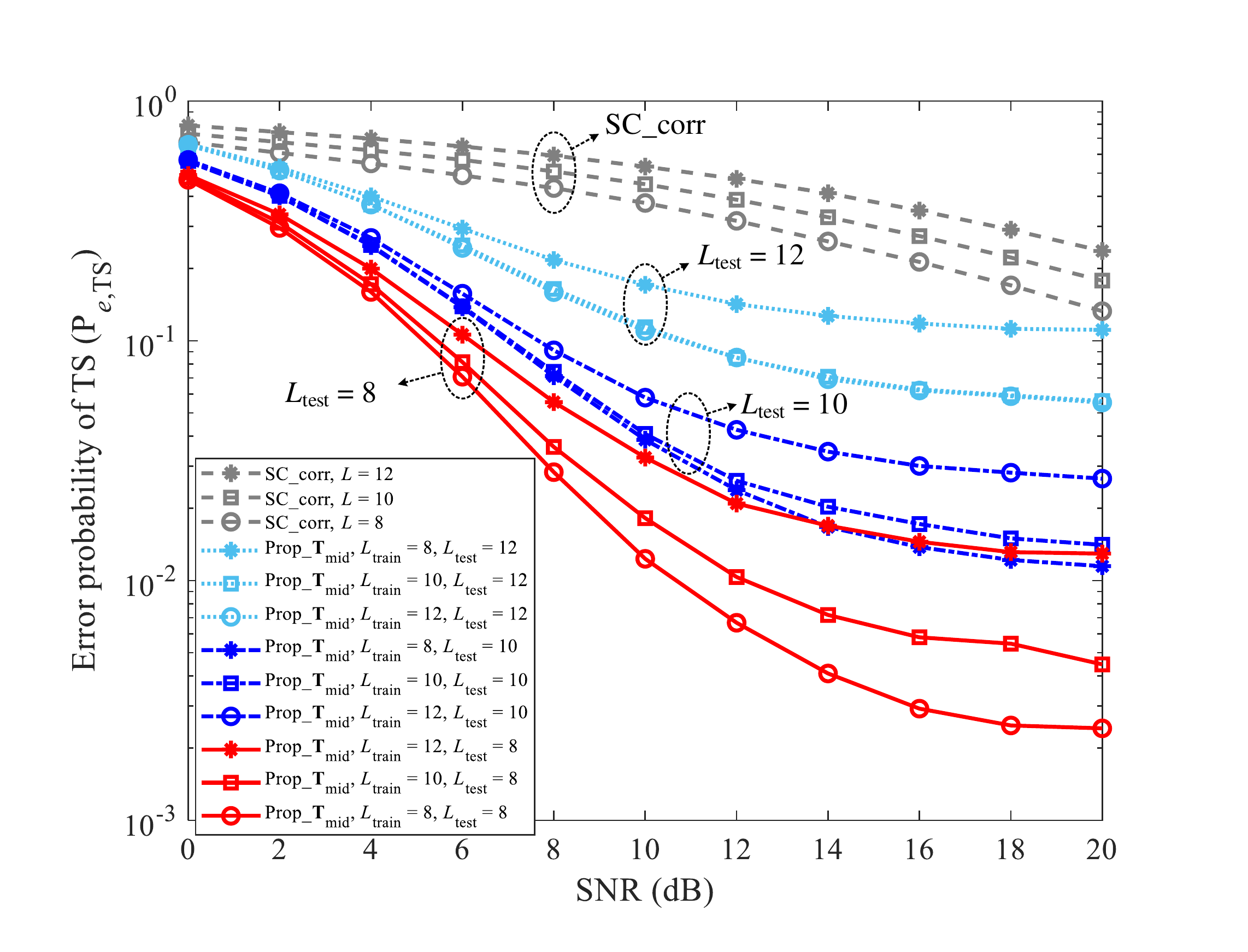}}
    \subfigure[``Prop\_$\textrm{T}_{\textrm{ISI-free}}$'']{
    \label{figL2_General}
    \includegraphics[scale=0.35]{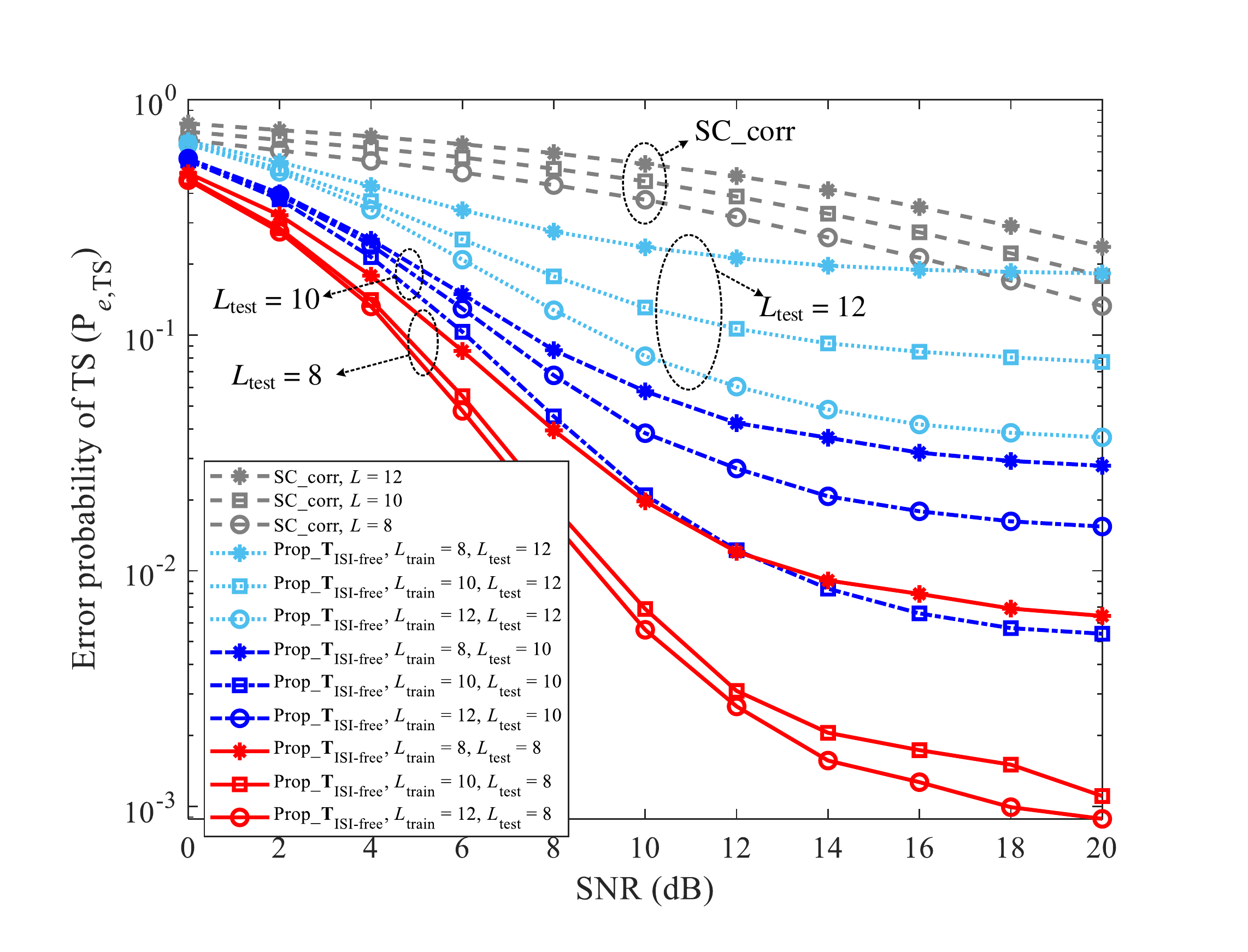}}
\caption{Generation performance against the impacts of $L$.}
\label{Sim2}
\end{figure}
\begin{figure}[h]
\centering
    \subfigure[``Prop\_$\textrm{T}_{\textrm{mid}}$'']{
    \label{figL1_General}
    \includegraphics[scale=0.35]{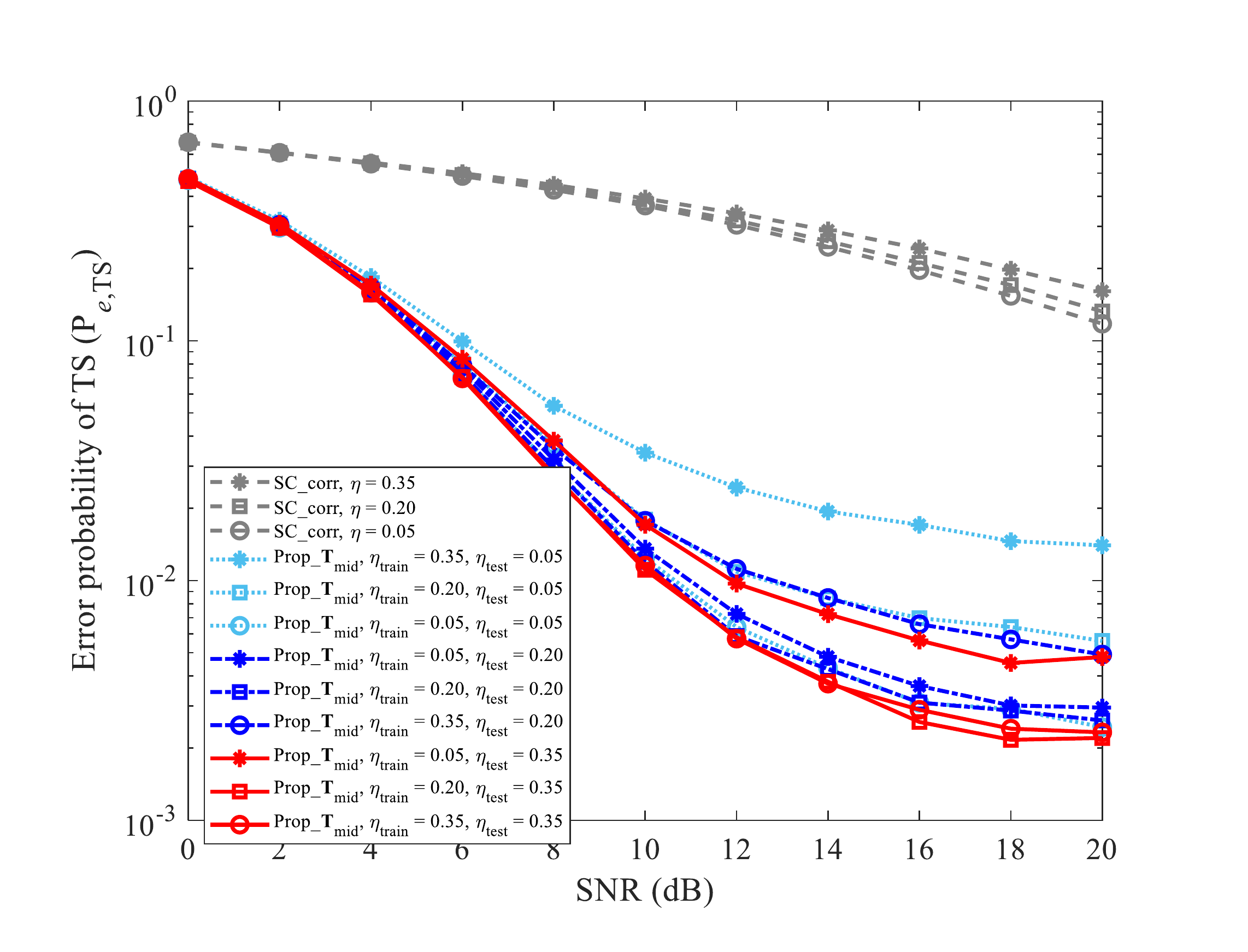}}
    \subfigure[``Prop\_$\textrm{T}_{\textrm{ISI-free}}$'']{
    \label{figL2_General}
    \includegraphics[scale=0.35]{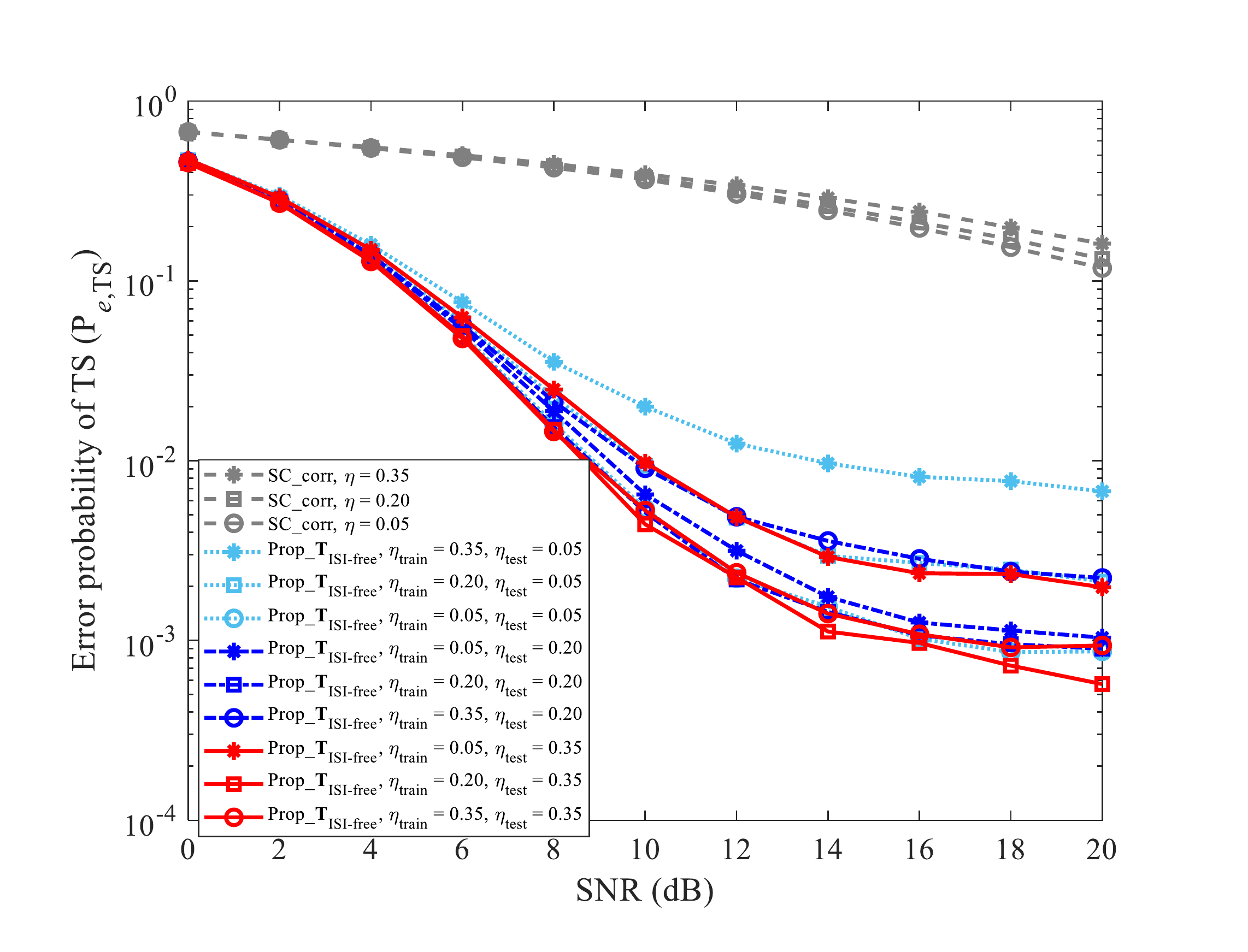}}
\caption{Generation performance against the impacts of $\eta$.}
\label{Sim3}
\end{figure}
\subsection{Generalization Analysis}
Fig.~\ref{Sim2} and Fig.~\ref{Sim3} plot the error probability of TS against the impacts of $L$ and $\eta$, respectively. For expression convenience, the subscript ``${}_{\textrm{train}}$'' and ``${}_{\textrm{test}}$'' are used to distinguish between training and testing values of $L$ (or $\eta$). Except for the parameters that need to discuss, other parameters remain the same as those in Fig.~\ref{Sim1}.
\subsubsection{Generalization against L} Fig.~\ref{Sim2} gives the error probability of TS to generalization against the impacts of $L$. In this simulation, $L_{\textrm{train}}=8$, $L_{\textrm{train}}=10$, and $L_{\textrm{train}}=12$ are employed for training phase, while different testing values of $L_{\textrm{test}}$ (i.e., $L_{\textrm{test}} = 8$, $L_{\textrm{test}} = 10$, and $L_{\textrm{test}} = 12$) are considered for each value of $L_{\textrm{train}}$. From Fig.~\ref{Sim2}(a) to Fig.~\ref{Sim2}(b), for each case of $L_{\textrm{test}}$, the error probabilities of ``{Prop\_$\textrm{T}_{\textrm{mid}}$}'' and ``{Prop\_$\textrm{T}_{\textrm{ISI-free}}$}'' are lower than that of ``SC\_corr''. It is also worth noting that the error probabilities of ``{Prop\_$\textrm{T}_{\textrm{mid}}$}'' and ``{Prop\_$\textrm{T}_{\textrm{ISI-free}}$}'' increase with the enlarged difference between $L_{\textrm{test}}$ and $ L_{\textrm{train}}$. Especially, Fig.~\ref{Sim2}(b) reveals that the generalization of ``{Prop\_$\textrm{T}_{\textrm{ISI-free}}$}'' is worse than that of ``{Prop\_$\textrm{T}_{\textrm{mid}}$}''. Although the timing-error-probability performance of ``{Prop\_$\textrm{T}_{\textrm{ISI-free}}$}'' degrades obviously when $L_{\textrm{test}}-L_{\textrm{train}}$ is enlarged, ``{Prop\_$\textrm{T}_{\textrm{ISI-free}}$}'' still reaches the smallest error probability in most cases. To sum up, ``{Prop\_$\textrm{T}_{\textrm{mid}}$}'' and ``{Prop\_$\textrm{T}_{\textrm{ISI-free}}$}'' possess a relatively good TS performance when $L_{\textrm{test}}\neq L_{\textrm{train}}$, yet further improvement on the generalization of ``{Prop\_$\textrm{T}_{\textrm{ISI-free}}$}'' is needed.
\subsubsection{Generalization against $\eta$} To test the generalization of the proposed TS scheme against the impacts of $\eta$, Fig.~\ref{Sim3} gives the error probability of TS. In this simulation, $\eta_{\textrm{train}}=0.05$, $\eta_{\textrm{train}}=0.2$, and $\eta_{\textrm{train}}=0.35$ are employed for training phase, while different testing values of $\eta_{\textrm{test}}$ (i.e., $\eta_{\textrm{test}} = 0.05$, $\eta_{\textrm{test}} = 0.2$, and $\eta_{\textrm{test}} = 0.35$) are considered for each value of $\eta_{\textrm{train}}$. From Fig.~\ref{Sim3}(a) to Fig.~\ref{Sim3}(b), it could be observed that, the error probabilities of ``{Prop\_$\textrm{T}_{\textrm{mid}}$}'' and ``{Prop\_$\textrm{T}_{\textrm{ISI-free}}$}'' are smaller than that of ``SC\_corr'', which indicates the effectiveness of ``{Prop\_$\textrm{T}_{\textrm{mid}}$}'' and ``{Prop\_$\textrm{T}_{\textrm{ISI-free}}$}'' in the reduction of TS error probability, even for $\eta_{\textrm{test}}\neq\eta_{\textrm{train}}$.
For ``{Prop\_$\textrm{T}_{\textrm{mid}}$}'' and ``{Prop\_$\textrm{T}_{\textrm{ISI-free}}$}'', the variation of TS error probability is enlarged with the enlarged difference between $\eta_{\textrm{test}}$ and $ \eta_{\textrm{train}}$, but this influence of varying $\eta$ is not obvious on the error probabilities of ``Prop\_$\textrm{T}_{\textrm{mid}}$'' and ``Prop\_$\textrm{T}_{\textrm{ISI-free}}$''. Namely, although the generalization performances of ``{Prop\_$\textrm{T}_{\textrm{ISI-free}}$}'' and ``{Prop\_$\textrm{T}_{\textrm{ISI-free}}$}'' are relatively degraded when $\eta_{\textrm{test}}\neq\eta_{\textrm{train}}$, this two types of proposed schemes still reach the lower error probability than ``SC\_corr'' does. As a result, ``{Prop\_$\textrm{T}_{\textrm{mid}}$}'' and ``{Prop\_$\textrm{T}_{\textrm{ISI-free}}$}'' possess a good TS performance against $\eta$ when the training $\eta$ is not the testing $\eta$.
\section{Conclusion}
In this paper, an ELM-based TS scheme is proposed for the TS in OFDM system with nonlinear distortion. For the task of TS in an ELM-based network, the trained ELM model with timing preprocessing achieves a far smaller TS error probability than that without timing preprocessing. Meanwhile, with the ELM network used in our TS scheme, not only the nonlinear distortion is suppressed, but also the TM is refined. Especially, two types of novel labels exploiting the prior information of ISI-free regions are investigated. According to the analysis and simulations, the proposed ELM-based TS scheme presents the effectiveness in the reduction of TS error probability, and reveals its generalization for the cases where the training and testing channels are of different parameters of $L$ (or $\eta$).
\section{Acknowledge}
This work is supported in part by the National Key Research and Development Program (Grant No. 2018YFB1800800), the Sichuan Science and Technology Program (Grant No. 2021JDRC0003), the Major Special Funds of Science and Technology of Sichuan Science and Technology Plan Project (Grant No. 19ZDZX0016 /2019YFG0395), the Demonstration Project of Chengdu Major Science and Technology Application (Grant No. 2020-YF09- 00048-SN), and the Special Funds of Civil-Military Integration Industry Development of Sichuan Province (Grant No. zyf-2018-056).




%\bibliographystyle{IEEEtran}
%\nocite{*}
%\bibliographystyle{ieeetr}
%\bibliography{ref}

%\begin{thebibliography}{1}
%\end{thebibliography}

\end{document}